\documentclass[a4paper]{article}

\IfFileExists{xurl.sty}{\usepackage{xurl}}{} 

\usepackage{amsmath, amsthm, amsfonts,amssymb}

\usepackage[utf8]{inputenc}

\usepackage{graphicx}
\usepackage{relsize}
\usepackage{natbib}
\usepackage[noblocks]{authblk}

\usepackage[margin=3.5cm]{geometry}

\usepackage[usenames,dvipsnames]{color}

\usepackage{xspace}
\usepackage{multirow, array}
\usepackage{longtable}
\newtheorem{prop}{Proposition}[section]

\newtheorem{defn}{Definition}[section]

\usepackage{array}
\newcolumntype{x}[1]{>{\centering\arraybackslash\hspace{0pt}}p{#1}}

\newcommand{\R}{\mathbb{R}}

\newcommand{\SI}{\mathcal{SI}}
\newcommand{\SIW}{\mathcal{SIW}}

\newcommand{\G}{\mathcal{G}}
\newcommand{\SG}{\mathcal{SG}}
\newcommand{\Pa}{\mathcal{P}}
\newcommand{\M}{\mathcal{M}}

\newcommand{\f}{\mathsf{f}}
\newcommand{\DP}{\mathsf{DP}}
\newcommand{\PG}{\mathsf{PG}}

\author[1,5]{J.M. Alonso-Meijide}
\author[2,3,5]{L.M. Armijos-Toro}
\author[3,5]{B. Casas-Méndez}
\author[4,5]{M.A. Mosquera}

\affil[1]{MODESTYA Research Group, Department of Statistics, Mathematical Analysis and Optimisation, Faculty of Sciences, University of Santiago de Compostela, Campus de Lugo, 27002 Lugo, Spain.}
\affil[2]{Departamento de Ciencias Exactas, Universidad de las Fuerzas Armadas ESPE, Sangolquí, Ecuador.}
\affil[3]{MODESTYA Research Group, Department of Statistics, Mathematical Analysis and Optimisation and IMAT, Faculty of Mathematics, University of Santiago de Compostela, Campus Vida, 15782 Santiago de Compostela, Spain.}
\affil[4]{Universidade de Vigo, Departamento de Estat\'{i}stica e Investigaci\'{o}n Operativa, 32004 Ourense, Spain.}
\affil[5]{CITMAga, 15782 Santiago de Compostela, Spain.}

\date{}

\title{An application of power indices for the family of weighted majority games in partition function form}

\begin{document}
\maketitle

\begin{abstract}
Based on \cite{Holler1982} and \cite{Armijos2021} we propose two power indices to measure the influence of the players in the class of weighted majority games in partition function form. We compare these new power indices with their original versions on the class of games in characteristic function form. Finally, we use both pairs of power indices for games in partition function form to study the distribution of power in the National Assembly of Ecuador that emerged after the elections of February 7, 2021.

\vskip 3mm

\noindent \emph{Key words}: Weighted majority games, Partition function form games, Minimal winning embedded coalitions, Power indices, Colomer-Martínez power index, Public Good power index.

\vskip 3mm

\noindent \textit{Math. Subj. Class. (2020)}: 91A12, 91A80

\end{abstract}

\section{Introduction}

Game theory is a mathematical discipline that is dedicated to the study of decision problems in which various players interact. In a cooperative game the players have mechanisms that allow them to take binding agreements. One of the main lines of research in cooperative game theory is the study of simple games and power indices. In a simple game, the worth of a coalition is 1 or 0, and then, there are two types of coalitions: winning and losing. Simple games are supposed to be monotonic, in the sense that the enlargement of a winning coalition cannot cause to become it in a losing one. Frequently, simple games model voting situations. A power index provides a measure of a voter's ability to change the outcome of a voting. Many different power indices have been defined and studied. The most well known are the Shapley-Shubik power index (\citealp{Shapley1954}) and the Banzhaf power index (\citealp{Banzhaf1964}). Both power indices assign to a player a measure that is based on the contributions the player makes by joining other coalitions. A winning coalition is minimal if it becomes a losing one if any of its members cease to be part of it, that is, all the players of a minimal winning coalition are decisive. \cite{Deegan1978} and \cite{Holler1982} proposed two power indices that assign a measure of power to a player taking into account only the minimal winning coalitions to which he belongs: the Deegan-Packel power index and the Public Good power index, respectively. The Deegan-Packel power index is based on the assumption that all minimal winning coalitions are equally likely, and that within a coalition all its players are equally important. The Public Good power index does not take into account the number of players who are part of the minimal winning coalitions, but the cardinal of the set of minimal winning coalitions to which a player belongs. Other power indices that follow similar arguments are those proposed in \cite{Johnston1978} and \cite{Alvarez2015a}. The Johnston power index takes into account only those winning coalitions in which there are some decisive players. The power index proposed by \cite{Alvarez2015a} only takes into account winning coalitions that do not contain any null player.

A particular class of simple games is the family of weighted majority games. A simple game is a weighted majority if there is a set of weights for the players and a quota, such that a coalition is winning if and only if the sum of the weights of the players of the coalition (that is, the weight of the coalition) is not less than the quota. \cite{Colomer1995} proposed a power index that is specific for the family of weighted majority games. The Colomer-Martínez power index, like the Deegan-Packel or the Public Good power indices, is also based on minimal winning coalitions. This power index has in common with the Deegan-Packel power index that all the minimal winning coalitions are considered equally likely, but it does not consider that all the players are equally important. The relevance of the players of a minimal winning coalition is directly related with their weights. \cite{Barua2005} defined a different power index for the family of weighted majority games. Although this power index is based on the Banzhaf power index, when absolute majority is required, the power index defined in Barua et al. (2005) assigns power in a proportional way to the weight of each player, analogous to the way the Colomer-Martínez power index acts. In \cite{Armijos2021}, a new power index for the family of weighted majority game was defined and axiomatically characterized. This new power index assigns to a player an amount that is defined considering ideas of both the Public Good and Colomer-Martínez power indices, because it takes into account:  i) the cardinal of the minimal winning coalitions to which he belongs and ii) his weight in the majority game.

In addition to the simple games discussed above, two other relevant types of cooperative games are as follows: games in characteristic function form and games in partition function form. In a game in characteristic function form a coalition gets an amount regardless of how the rest of the players organize. The characteristic function assigns a value to each subset of players. Note that simple games are a subclass of the games in characteristic function form. However, in the more general model of games in partition function form (\citealp{Thrall1963}), what a coalition gets depends on the arrangement of the rest of the players. The partition function assigns a value to each embedded coalition, that is, to each pair formed by a coalition (called active coalition) and a partition of the players outside the active coalition. Returning to the voting situations modeled by the simple games, the model of game in partition function form has proven to be useful to represent situations with plurality rules. In these cases, if no candidate is supported by a qualified majority, that which get more votes get the government. Thus, in a weighted majority game in partition function form, an embedded coalition is considered winning if the weight of the active coalition is greater than or equal to the weight of any of the coalitions that constitute the partition of the rest of the players. In games in characteristic function form, a solution concept is intended to distribute the value of the total coalition among all players and two of the most prominent solution concepts are the so-called Shapley (\citealp{Shapley1953}) and Banzhaf values, respectively. \cite{Myerson1977} studied a generalization of the Shapley value for games in partition function form. \cite{Dutta2010} studied a family of values that generalizes the Shapley value. \cite{Clippel2008} proposed the externality-free Shapley value. \cite{Bolger1983} and \cite{Bolger1990} studied generalizations of the Banzhaf values for games in partition function form. In \cite{Alvarez2017}, the restriction of the externality-free Shapley value (the externality-free Shapley-Shubik power index) was axiomatically characterized. In \cite{Alvarez2015}, an alternative generalization of the Banzhaf value (the so called ordinal Banzhaf power index) was proposed for the family of simple games in partition function form.  

In \cite{Alonso2017} a new class of simple games in partition function form was defined. Moreover, generalizations of the Deegan-Packel and Public Good power indices for these simple games in partition function form are proposed and characterized. To define these power indices, a notion of inclusion among embedded coalitions and a related property of monotonicity are introduced. Roughly speaking, an embedded coalition is larger when the active coalition does not decrease and the partition of the rest of players becomes finer. \cite{Carreras2008} used a similar idea even though in a different setting. 

There is also a large body of literature in which the methodology described above is applied to real problems, particularly in the field of politics, to measure the power of the different political parties represented in parliaments or other organizations. In \cite{Alonso2017}, the new power indices for simple games in partition function form there introduced, were used to study the distribution of power in the Parliament of Andalusia, an autonomous community in Spain, that emerged after the elections of March 22, 2015. In \cite{Arevalo2020}, some other well-known power indices extended to simple games in partition function form were applied to the Parliament of other autonomous community in Spain, the Basque Country. \cite{Armijos2021} presented an application of their new power index for the family of weighted majority games to the results of the elections held in 2021 for the formation of a new National Assembly of Ecuador. In this last case, the so-called legislative benches, also known as blocks, are the official political groupings within the Assembly, with the right to have authority within the different legislative commissions. If initially, after the elections, there is a configuration of these blocks, it may happen that several additional parliamentary groups are formed during the course of the legislative period, due to splits within the legislative benches, which raises the interest in analyzing the evolution of power within this Assembly over the course of a single legislative period. 

In this paper, the Colomer-Martínez power index and the power index proposed in \cite{Armijos2021}, are generalized for the model of weighted majority games in partition function form. They are also applied to the analysis of power within the National Assembly of Ecuador. The rest of the paper is organized as follows. In Section 2 we introduce some preliminaries and the basic notation for weighted majority games in partition function form. In Section 3, we generalize to this context of games in partition function form the Colomer-Martínez power index and the power index proposed in \cite{Armijos2021}. Finally, in Section 4 a real example from the political field, the National Assembly of Ecuador, is used to illustrate the new power indices. To do this, we analyze minimal winning embedded coalitions in partition function form (plurality rule). To complete our study, we compute several power indices from the literature and compare the results.

\section{Simple games in characteristic and partition function form}

In this section we introduce the notation and preliminary concepts necessary for our study focusing on weighted majority games in partition function form.

\subsection{Simple games in characteristic function form}

A game in characteristic function form is a pair $(N,v)$ where $N$ is the finite set of players and $v: 2^N \longrightarrow\R $ is the characteristic function of the game satisfying $v(\emptyset)=0$. Non-empty subsets of $N$ are called coalitions. The number $v(S)$ is to be understood as the worth of coalition $S$ regardless of how the player of $N\setminus S$ organize. A game $(N,v)$ is monotone when $v(S)\leq v(T)$ for all $S\subseteq T \subseteq N$. A simple game is a monotone game such that $v(N) = 1$ and, for every $S \subseteq N$, $v(S) = 0$ or $v(S) =1$. We denote by $\SI^N$ the class of simple games with set of players $N$ and by $\SI$ the class of simple game with an arbitrary set of players. Let $(N,v)\in \SI^N$ then, a coalition $S\subseteq N$ is a winning (resp. losing) coalition when $v(S)=1$ ($v(S)=0$). $W(v)$ denotes the set of winning coalitions for the game $(N,v)$. A simple game $(N,v)$ is called decisive whenever $S\in W(v)$ if, and only if, $N\setminus S \notin W(v)$ for each $S\subseteq N$. A winning coalition $S \subseteq N$ is a minimal winning coalition if there no exists $T\subset S$ such that $T$ is a winning coalition. We denote by $M(v)$ the set of minimal winning coalitions of the game $(N,v)$ and let $M_{i}(v) = \{S \in M(v) : i \in S \}$ for each $i\in N$. 

Let $S\subseteq N$, $S\neq \emptyset$. The game $(N,u_S) \in \SI^N$ (called unanimity game of $S$) is the (non-decisive) simple game such that $M(u_S)=\{S\}$, that is, for every $T\subseteq N$,

\begin{equation}\label{eq:basis-SIN}
u_S(T)=\begin{cases}
1 & \text{if } S\subseteq T\\
	0 & \text{otherwise}.
\end{cases}
\end{equation}

It is well-known that $\left\{(N,u_S) : S\in 2^N\backslash \emptyset\right\}$ constitutes a basis of the vector space of games in characteristic function form with set of players $N$.

Let $(N,v), (N,v') \in  \SI^N$. Their union is the game $(N,v \vee v') \in \SI^N$ such that, for all $S \subseteq N$, $S \in W(v \vee v')$ if $S \in W(v)$ or $S \in W(v')$; or in an equivalent way, $(v \vee v')(S) = \max\{v(S),v'(S)\}$, for every $S\subseteq N$. Two games $(N,v), (N,v') \in  \SI^N$ are mergeable if for all pair of coalitions $S,T \subseteq N$ such that $S \in M(v)$ and $T \in M(v')$, it holds that $S \nsubseteq T$ and $T \nsubseteq S$. If $(N,v), (N,v') \in \SI^N$ are mergeable, then $M(v)\cap M(v')=\emptyset$ and $M(v\vee v')=M(v)\cup M(v')$.

A power index for the family of simple games in characteristic function form is a mapping, $f$, that assigns to every simple game $(N,v)\in \SI$ a vector $f(N,v)\in \R^N$, where $f_i(N,v)$ describes the power of agent $i$ in the game $(N,v)$. Throughout this paper, we only consider efficient power indices, that is, $\sum_{i \in N} f_i(N,v)=1.$

The definitions of two power indices, the Deegan-Packel power index (\citealp{Deegan1978}) and the Public Good power index (\citealp{Holler1982}) that have in common that only take into account the minimal winning coalitions of the game are presented below.

\begin{defn}
The Deegan-Packel power index ($DP$) is the power index defined for every $(N,v)\in \SI$ and $i\in N$ by $$DP_i(N,v)=\frac{1}{|M(v)|}\sum_{S \in M_i(v)}\frac{1}{|S|}.$$
\end{defn}

\begin{defn}
The Public Good power index ($PG$) is the power index defined for every $(N,v)\in \SI$ and $i\in N$ by $$PG_i(N,v)=\frac{|M_i(v)|}{\sum\limits_{j\in N}|M_j(v)|}.$$
\end{defn}

Although both power indices use only minimal winning coalitions, for the computation of the $DP$ power index the number of players that are part of each minimal winning coalition is taken into account, while for the $PG$ power index only the number of minimal winning coalitions to which a player belongs is used, regardless of their size.

A null player in a game $(N,v) \in \SI$ is a player $i \in N$ such that $M_{i}(v)= \emptyset$. Two players $i,j \in N$ are symmetric in a game $(N,v) \in \SI$ if, for all coalition $S \subseteq N \backslash \{i,j\}$ such that $S \notin W(v)$, $S \cup \{i\} \in W(v)$ if and only if $S \cup \{j\} \in W(v)$. A power index satisfies the property of null player if $f_i(N,v)=0$ when $i$ is a null player in $(N,v)$. A power index satisfies the property of symmetry if $f_i(N,v)=f_j(N,v)$ when $i,j \in N$ are symmetric in $(N,v)$. Both power indices, $DP$ and $PG$ satisfy null player and symmetry properties. Moreover, $DP$ and $PG$ power indices of a union game $(N,v \vee v')$ of two mergeable games $(N,v)$ and $(N,v')$ can be computed as a weighted sum of the corresponding power indices of the games $(N,v)$ and $(N,v')$, with different weights in each case (see \citealp{Deegan1978,Holler1983}). For every pair of mergeable simple games $(N,v),(N,v') \in \SI^N$ it holds that:

$$DP_i(N,v \vee v') = \frac{|M(v)| DP_i(N,v) +|M(v')| DP_i(N,v')}{|M(v \vee v')|}\ {\rm and}$$ 

$$PG_i(N,v \vee v') =\frac{\sum_{j \in N} |M_{j}(v)| PG_i(N,v) +\sum_{j \in N} |M_{j}(v')| PG_i(N,v')}{\sum_{j \in N} |M_{j}(v \vee v')|}.$$

In this paper, we study the family of weighted majority games, a particular class of simple games. 
\begin{defn}

A simple game $(N,v) \in \SI$ is a weighted majority game if there exist a non-negative vector of weights, $\boldsymbol{w}=(w_{1},w_{2},...,w_{n})$, where $w_i$ is the weight of player $i\in N$, and a quota $q > 0$ such that a coalition $S \in W(v)$ if and only if $w(S) = \sum_{i \in S} w_{i} \geq q$. 

\end{defn}

We denote by $\SIW^N$ the class of weighted majority games with set of players $N$ and by $\SIW$ the class of simple game with an arbitrary set of players. We identify a weighted majority game by the tuple $[q; w_{1}, w_{2}, ..., w_{n}]$, or $[q;\boldsymbol{w}]$ when no confusion is possible. A majority game is a weighted majority game such that $w_i=1$, for all $i \in N$. If $n$ is an odd number then the majority game $[(n+1)/2;1,...,1]$ is decisive.

\cite{Colomer1995} defined a power index for the family of weighted majority games. This power index could be considered as a non-symmetric generalization of the Deegan-Packel power index.

\begin{defn}
The Colomer-Martínez power index (CM) is the power index defined for every $[q; \boldsymbol{w}]\in \SIW$ and $i\in N$ by $$CM_i(q;\boldsymbol{w})=\frac{1}{|M(q;\boldsymbol{w})|}\sum_{S \in M_i(q;\boldsymbol{w})}\frac{w_i}{\sum_{j \in S}w_j}.$$
\end{defn}

In \cite{Armijos2021} the $HCM$ power index, a new power index for weighted majority games, was proposed. The definition of the $HCM$ power index shares ideas with the Colomer-Martínez and Public Good power indices. Like the Public Good power index, the total number of minimal winning coalitions to which a player belongs is considered, and like the Colomer-Martínez power index, his weight is taking into account to define the power index.

\begin{defn}
The Holler-Colomer-Martínez power index is the power index defined for every $[q; \boldsymbol{w}]\in \SIW$ and $i\in N$ by $$HCM_i(q;\boldsymbol{w}) =  \frac{|M_{i}(q;\boldsymbol{w})|w_i}{\sum_{j \in N} |M_{j}(q;\boldsymbol{w})|w_j }.$$
\end{defn}	
The $HCM$ power index computes the power of each player as the proportion, balanced by its weights in the game, of the number of minimal winning coalitions to which he belongs. An alternative expression for the $HCM$ power index of a player $i \in N$ in a simple game $[q; \boldsymbol{w}]\in \SIW^N$ is

$${HCM}_{i}(q;\boldsymbol{w}) = \frac{\sum_{S \in M_{i}(q;\boldsymbol{w})} w_{i}}{\sum_{j \in N} \sum_{S \in M_{j}(q;\boldsymbol{w})} w_{j}} = \sum_{S \in M_{i}(q;\boldsymbol{w})}\frac{ w_{i}}{\sum_{T \in M(q;\boldsymbol{w})} \sum_{j\in T} w_{j}}.$$

When the Deegan-Packel and the Public Good power indices are restricted to the family of weighted majority games, the main difference between these indices and the Colomer-Martínez and the $HCM$ power indices is that the last two ones are not symmetric. In a simple game with a unique minimal winning coalition $S$, the Deegan-Packel and the Public Good power indices assign the same power to all players in $S$, but the Colomer-Martínez and the $HCM$ power indices assign to each player of $S$ a power directly related with his weight. Formally, given $(N,v)\in \SIW^N$, determined by $[q; \boldsymbol{w}]$, and $S\in 2^N\backslash \emptyset$ with $M(v)=S$, then $DP_i(N,v)=PG_i(N,v)=1/|S|$ for every $i \in S$, meanwhile $CM_i(q;\boldsymbol{w})=HCM_i(q;\boldsymbol{w})={w_i}/{\sum_{j \in S}w_j}$. Moreover, for every $i,j \in S$, ($i$ and $j$ are symmetric),
$$DP_i(N,v)=PG_i(N,v)=DP_j(N,v)=PG_j(N,v),$$ $$CM_i(q;\boldsymbol{w})w_j=CM_j(q;\boldsymbol{w})w_i=HCM_i(q;\boldsymbol{w})w_j=HCM_j(q;\boldsymbol{w})w_i.$$
So, the following property is satisfied by the Colomer-Martínez and the $HCM$ power indices. A power index on $\SIW$ satisfies the property of weighted symmetry if $w_jf_i(q;\boldsymbol{w})=w_if_j(q;\boldsymbol{w})$ when $i,j \in S$ and $M(q;\boldsymbol{w})=\{S\}$. Moreover, Colomer-Martínez and the $HCM$ power indices also satisfy some merging properties adapted to the class of weighted majority games (see \citealp{Armijos2021}).

\subsection{Simple games in partition function form}

We denote by $\Pa(N)$ the set of partitions of a finite set $N$. We assume that the empty set is an element of every partition, that is, $\emptyset\in \mathsf{P}$ for every $\mathsf{P}\in\Pa(N)$. A partition $\mathsf{P}$ is coarser than $\mathsf{Q}$ if every block $B \in \mathsf{Q}$ is included in some block $A \in \mathsf{P}$, i.e. $\mathsf{P}\supseteq \mathsf{Q}$. Then $(\Pa(N),\supseteq)$ is a lattice, called the partition lattice. With this ordering, the bottom element of the lattice is the finest partition $\{\{1\}, . . . , \{n\}\}$, while the top element is the coarsest partition $\{N\}$. An embedded coalition of $N$ is a pair $(S;\mathsf{P})$ where $\mathsf{P}\in\Pa(N)$ and $S\in \mathsf{P}$, is the active coalition in $\mathsf{P}$. We denote by $EC^N$ the set of embedded coalitions of $N$, i.e. $EC^N=\{(S;\mathsf{P}) : \mathsf{P}\in\Pa(N) \mbox{ and }S\in \mathsf{P}\}$. We say that a player $i\in N$ participates in an embedded coalition $(S;\mathsf{P})\in EC^N$ if player $i$ belongs to $S$. We simplify and write $S\cup i$ and $S\setminus i$ instead $S\cup \{i\}$ and $S\setminus \{i\}$, respectively. Given $\mathsf{P}\in\Pa(N)$ and $i\in N$, we denote by $\mathsf{P}(i)$ to the element of $\mathsf{P}$ that contains $i$, i.e., $\mathsf{P}(i)\in \mathsf{P}$ and $i\in \mathsf{P}(i)$.

A game in partition function form is a pair $(N,v)$, where $N$ is the finite set of players and $v:EC^N\rightarrow \R$ is the partition function of the game satisfying $v(\emptyset;\mathsf{P})=0$ for every $\mathsf{P}\in\Pa(N)$. The number $v(S;\mathsf{P})$ is to be understood as the worth of coalition $S$ when the players are organized according to $\mathsf{P}$. In a game in partition function form it is possible that $v(S;\mathsf{P}) \not =v(S;\mathsf{P}')$, for two pair of partitions $\mathsf{P},\mathsf{P}' \in\Pa(N)$ with $S\in \mathsf{P}$ and $S\in \mathsf{P}'$ . The set of games in partition function form with common set of players $N$ is denoted by $\G^N$ and the set of games in partition function form with an arbitrary set of players is denoted by $\G$. Is is easy to notice that $\G^N$ is a vector space over $\R$. Indeed, \cite{Clippel2008} devised a basis of the vector space that generalizes the basis of games in characteristic function form that consists of unanimity games defined in Eq. (\ref{eq:basis-SIN}). Given $(S;\mathsf{P})\in EC^N$, with $S\neq \emptyset$, let $\left(N,e_{(S;\mathsf{P})}\right)\in\G$ be defined for every $(T;\mathsf{Q})\in EC^N$ by
\begin{equation}\label{eq:basis-GN}
e_{(S;\mathsf{P})}(T;\mathsf{Q})=\begin{cases}1 & \mbox{if }S\subseteq T \mbox{ and }\forall T'\in \mathsf{Q}_{-T}, \exists S'\in \mathsf{P} \mbox{ such that } T'\subseteq S' \\ 0 &\mbox{otherwise,}\end{cases}
\end{equation}
where $\mathsf{Q}_{-T}\in \Pa(N\backslash T)$ denotes the partition $\mathsf{Q}\backslash \{T\}$. \cite{Clippel2008} showed that $\left\{\left(N,e_{(S;\mathsf{P})}\right) : (S;\mathsf{P})\in EC^N \mbox{ and } S\neq \emptyset\right\}$ constitutes a basis of $\G^N$.

In this paper we are concerned with a subclass of $\G$ that generalizes simple games in characteristic function form as introduced by \cite{Neumann1944}. For doing so, \cite{Alonso2017} developed a concept of monotonicity for games in partition function form. The intuition behind monotonic games is that the enlargement of a coalition cannot cause a decrease in its worth. Therefore, in order to generalize this idea, \cite{Alonso2017} used a notion of inclusion for embedded coalitions that will be of key importance for its results and that it is implicitly formulated in Eq. (\ref{eq:basis-GN}).

\begin{defn}\label{def:inclusion}
Let $N$ be a finite set and $(S;\mathsf{P}),(T;\mathsf{Q})\in EC^N$. We define the inclusion among embedded coalitions as follows: \begin{equation*}
(S;\mathsf{P})\sqsubseteq (T;\mathsf{Q}) \Longleftrightarrow S\subseteq T \mbox{ and }\forall T'\in \mathsf{Q}_{-T}, \exists S'\in \mathsf{P} \mbox{ such that } T'\subseteq S'.
\end{equation*}
\end{defn}

For instance, $$(\{1\};\{1\},\{2,3,4\})\sqsubseteq (\{1,2\};\{1,2\},\{3,4\})\ {\rm and}\ (\{1,2\};\{1,2\},\{3,4\})\sqsubseteq (\{1,2\};\{1,2\},\{3\},\{4\}).$$ Nevertheless $(\{1\};\{1\},\{2\},\{3\},\{4\})$ and  $(\{1,2\};\{1,2\},\{3,4\})$ are not comparable. Note that whenever $S\neq \emptyset$, $(S;\mathsf{P})\sqsubseteq (T;\mathsf{Q})$ if and only if $e_{(S;\mathsf{P})}(T;\mathsf{Q})=1$. According to the above definition, an embedded coalition $(S;\mathsf{P})$ is a subset of another embedded coalition $(T;\mathsf{Q})$ if $S\subseteq T$ and the partition of $N\setminus T$ defined as $\{R\setminus T : R\in \mathsf{P}\}$ is coarser than $\mathsf{Q}_{-T}$. 

We introduce the class of simple games in partition function form for which we first extend the notion of monotonicity to the games in partition function form. 

\begin{defn}\label{def:MON}
A game in partition function form $(N,v)\in\G$ is monotone when $v(S;\mathsf{P})\leq v(T;\mathsf{Q})$ for all $(S;\mathsf{P}), (T;\mathsf{Q})\in EC^N$ such that $(S;\mathsf{P})\sqsubseteq (T;\mathsf{Q})$.
\end{defn}

\begin{defn}\label{def:SG}
A game in partition function form $(N,v)\in\G$ is a simple game in partition function form if it satisfies:\begin{itemize}
\item[i)] $v(S;\mathsf{P})\in\{0,1\}$, for every $(S;\mathsf{P})\in EC^N$.\
\item[ii)] $v(N;\{\emptyset,N\})=1$.
\item[iii)] $(N,v)$ is monotone.
\end{itemize}
An embedded coalition, $(S;\mathsf{P})\in EC^N$, is winning if $v(S;\mathsf{P})=1$ and losing if $v(S;\mathsf{P})=0$. We denote by $\SG^N$ the set of simple games in partition function form with common set of players $N$ and by $\SG$ the set of simple games in partition function form with an arbitrary set of players.
\end{defn}

The simple games in partition function form, as defined above, are the generalization of simple games in characteristic function form form. First, each embedded coalition is either winning or losing. Second, the grand coalition $(N;\{\emptyset,N\})$  is always a winning coalition. Third, suppose that $(S;\mathsf{P})\in EC^N$ is a winning embedded coalition, then $(T;\mathsf{Q})$ is a winning embedded coalition when $(S;\mathsf{P})\sqsubseteq (T;\mathsf{Q})$, i.e., the game is monotone. The games that form the basis of \cite{Clippel2008}, see Eq. (\ref{eq:basis-GN}), are examples of simple games in partition function form.

In this paper, a particular class of embedded coalitions, the minimal winning embedded coalitions, play a very important role. Let $(N,v)\in\SG$. A winning embedded coalition, $(S;\mathsf{P})\in EC^N$ ($v(S;\mathsf{P})=1$) is a minimal winning embedded coalition if every proper subset of it is a losing embedded coalition, i.e., if $(T;\mathsf{Q})\sqsubset (S;\mathsf{P})$ implies that $v(T;\mathsf{Q})=0$.\footnote{A proper subset, $(T;\mathsf{Q})\sqsubset (S;\mathsf{P})$, is a subset $(T;\mathsf{Q})\sqsubseteq (S;\mathsf{P})$ satisfying $(T;\mathsf{Q})\neq(S;\mathsf{P})$.} The set of minimal winning embedded coalitions of a simple game $(N,v)$ in partition function form is denoted by $\M(v)$ and the subset of minimal winning embedded coalitions such that a given player $i\in N$ participates is denoted by $\M_i(v)$, i.e., $\M_i(v)=\left\{(S;\mathsf{P})\in\M(v) : i\in S\right\}$. Taking into account the inclusion relation among embedded coalition, a minimal winning embedded coalition $(S;\mathsf{P})\in EC^N$ is a winning embedded coalition such that the active coalition $S$ be of a minimum size and inactive coalitions $\mathsf{P}\setminus S$ to be of maximum size. 

A player $i\in N$ is a null player in $(N,v)\in \SG$ if he does not participate in any minimal winning embedded coalition, i.e., $\M_i(v)=\emptyset$. Two players $i,j\in N$ are symmetric in $(N,v)\in \SG$ if exchanging the two players does not change the type of a coalition, i.e., if for every $(S;\mathsf{P})\in EC^N$ such that $S\subseteq N\setminus \{i,j\}$,  $$v\left(S\cup i;\mathsf{P}_{-S,\mathsf{P}(i)}\cup\{S\cup i,\mathsf{P}(i)\setminus i\}\right)=1 \Leftrightarrow v\left(S\cup j;\mathsf{P}_{-S,\mathsf{P}(j)}\cup\{S\cup j,\mathsf{P}(j)\setminus j\}\right)=1,$$
where $\mathsf{P}_{-S,\mathsf{P}(k)}=\left(\mathsf{P}_{-S}\right)_{-\mathsf{P}(k)}$, for every $k\in \{i,j\}$.

In the same way that simple games in characteristic function form, a simple game in partition function form is completely determined by the set of minimal winning embedded coalitions. In a sense, all the relevant information of a simple game in partition function form is condensed in the set of minimal winning embedded coalitions. This fact is formally presented in the next result proved in \cite{Alonso2017}.

\begin{prop}\label{prop:min}
Let $\mathcal{C}\subseteq EC^N$ be such that there is no relation with respect to $\sqsubseteq $ between any pair  $(S;\mathsf{P}),(T;\mathsf{Q})\in \mathcal{C}$. Then, there exists a unique simple game in partition function form, $(N,v)$, such that $\M(v)=\mathcal{C}$.  
\end{prop}

\section{Power indices for weighted majority games in partition function form}

A power index for the family of simple games in partition function form is a mapping, $\f$, that assigns to every simple game in partition function form $(N,v)\in \SG$ a vector $\f(N,v)\in \R^N$, where $\f_i(N,v)$ describes the power of agent $i$ in the game $(N,v)$. 

In this paper, we are interested in the family of weighted majority games in partition function form, a particular class of simple games in partition function form. 

\begin{defn}\label{def:SGPFF}
A simple game  in partition function form $(N,v)\in \SG$ is a weighted majority game in partition function form if there exist a non-negative vector of weights, $\boldsymbol{w}=(w_{1},w_{2},...,w_{n})$, where $w_i$ is the weight of player $i\in N$, such that for every $(S;\mathsf{P})\in EC^N$
$$v(S;\mathsf{P})=1\quad \Longleftrightarrow \quad \sum_{i\in S}w_i\geq \sum_{i\in T}w_i \quad \forall T\in \mathsf{P}.$$
We denote by $\mathcal{SWG}^N$ the set of  weighted majority games in partition function form with set of players $N$ and by $\mathcal{SWG}$ the set of  weighted majority games in partition function form with an arbitrary set of players.
\end{defn}

In the following we extend the four power indices reviewed in Section 2.1 to this class of games.

\begin{defn}
$\DP$ is the power index defined for every $(N,v)\in\SG$ and $i\in N$ by $$\DP_i(N,v)=\frac{1}{|\M(v)|}\sum_{(S;\mathsf{P})\in\M_i(v)}\frac{1}{|S|}.$$
\end{defn}

\begin{defn}
$\PG$ is the power index defined for every $(N,v)\in\SG$ and $i\in N$ by $$\PG_i(N,v)=\frac{|\M_i(v)|}{\sum\limits_{j\in N}|\M_j(v)|}.$$
\end{defn}

The Colomer-Martínez and Holler-Colomer-Martínez power indices depend on the weights of each player, for this reason we identify a weighted majority game in partition function form $(N,v)$ by its vector of weights $\boldsymbol{w}$.

\begin{defn}
$\mathsf{CM}$ is the power index defined for every $(N,v)\in \mathcal{SWG}$, determined by the vector of weights $\boldsymbol{w}$, and $i\in N$ by $$\mathsf{CM}_i(\boldsymbol{w})=\frac{1}{|\M(\boldsymbol{w})|}\sum_{(S;\mathsf{P}) \in \M_i(\boldsymbol{w})}\frac{w_i}{\sum_{j \in S}w_j}.$$
\end{defn}

\begin{defn}
$\mathsf{HCM}$ is the power index defined for every $(N,v)\in \mathcal{SWG}$, determined by the vector of weights $\boldsymbol{w}$, and $i\in N$ by $$\mathsf{HCM}_i(\boldsymbol{w}) =  \frac{|\M_{i}(\boldsymbol{w})|w_i}{\sum_{j \in N} |\M_{j}(\boldsymbol{w})|w_j }.$$
\end{defn}

The two first extension were studied and characterized in \cite{Alonso2017}. The properties used in the characterizations are the natural generalizations of the properties used in the original versions of the Deegan-Packel and Public Good power indices (see \citealp{Deegan1978,Holler1982}) for simple games in characteristic function form. To the best of our knowledge, the last two extensions are presented for the first time in this article.

\section{A political example: The National Assembly of Ecuador}\label{sec:anda}

\begin{figure}[!h]
	\centering
	\caption{National Assembly of Ecuador in May 2021.}
	\vspace*{0.5cm}\includegraphics[width=0.7\linewidth]{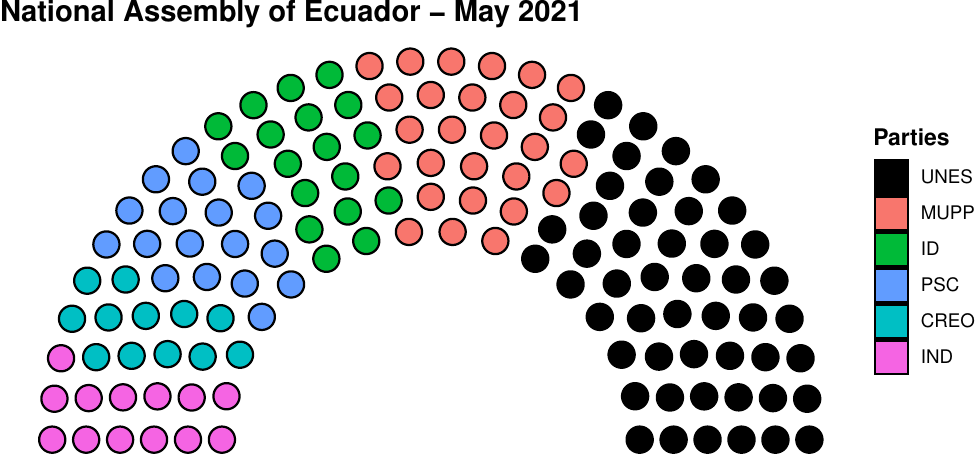} \label{Fig00}
\end{figure}

In this section, we evaluate the results of the proposed power indices for weighted majority games in partition function form applied to the National Assembly of Ecuador. The analysis of this Assembly using weighted majority games in characteristic function form can be found in \cite{Armijos2021}. The National Assembly of Ecuador consists of 137 assembly members. In February 2021, general elections were held in Ecuador\footnote{\url{https://www.primicias.ec/noticias/politica/los-cambios-en-las-bancadas-de-la-asamblea/}, last accessed 23/12/2021.}. The National Assembly was composed by: (49) UNES, (27) MUPP, (18) ID, (18) PSC, (12) CREO, and minorities (IND): (2) AVA, (2) MEU, (2) AH, (1) PSP, (1) AU, (1) MAP, (1) MUE, (1) MMI, (1) MAE, and (1) DEMSI.  The parties CREO and PSC are pro free market. Meanwhile, the parties ID and UNES are progressive trend. Finally, the party MUPP is closer to socialism. The current president of Ecuador belongs to CREO party. The results of assembly members together with the votes obtained by each party are summarized in Table \ref{Tb01} and Figure \ref{Fig00}.

\begin{table}
	\begin{center}
		\caption{National Assembly of Ecuador in May 2021.}
		\label{Tb01}
		\small{
			\begin{tabular}{ r|| c|c }
				\hline
				Parties & Votes     & Assembly members \\ \hline\hline
				UNES    & 5060922 &        49        \\
				MUPP    & 2530803 &        27        \\
				ID      & 1808867 &        18        \\
				PSC     & 1615833 &        18        \\
				CREO    & 1509436 &        12        \\
				IND     & 2061845 &        13        \\ \hline
			\end{tabular}}	
	\end{center}

\end{table}

\begin{figure}[!h]
	\centering
	\caption{Composition of the legislative benches - National Assembly of Ecuador in June 2021.}
	\includegraphics[scale=0.8]{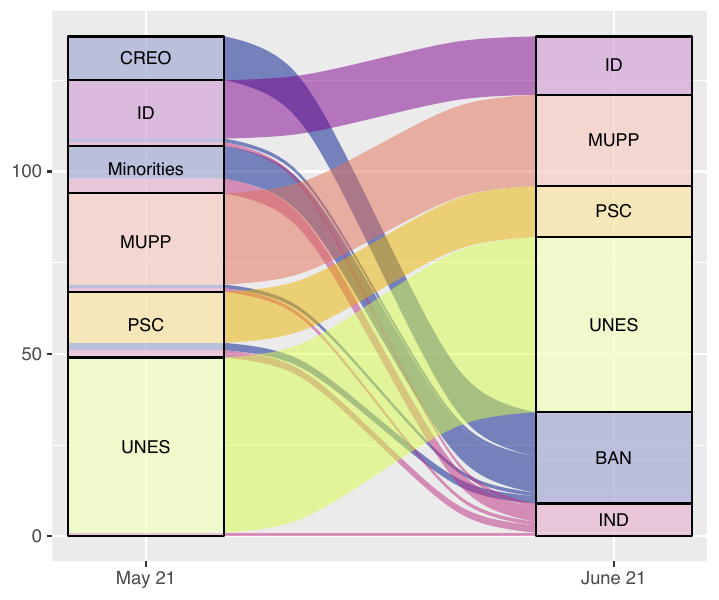} 
	\label{Fig01}
\end{figure}

In June 2021, the party of the president of Ecuador (CREO) and some assembly members of the other parties, mainly from minorities, consolidated the new legislative bench (25) BAN.  Also, some assembly members declared themselves independent: (9) IND. Figure \ref{Fig01} shows the redistribution of assembly members in the legislative benches in June 2021.

\begin{figure}[!h]
	\caption{National Assembly of Ecuador from June to December 2021.}\label{Fig10}
	\vspace*{0.5cm}
	\begin{center}
	\begin{tabular}{cc}
		\includegraphics[width=0.45\linewidth]{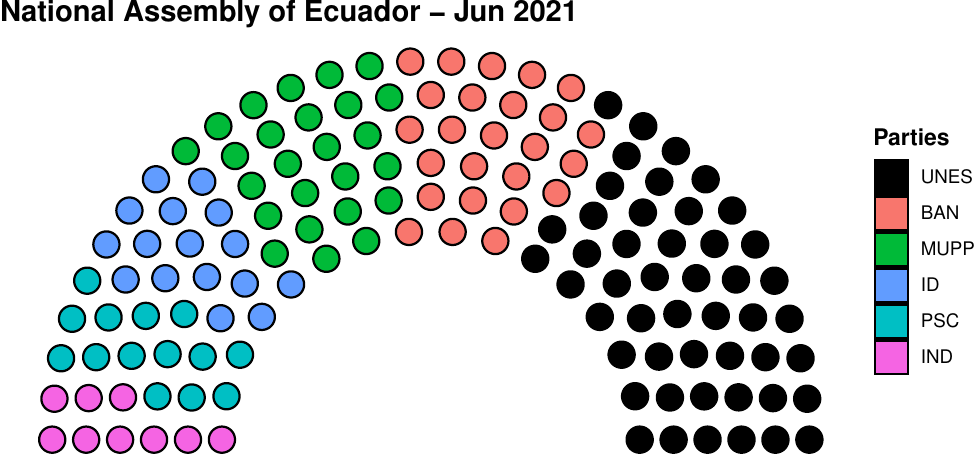} &
		
		\includegraphics[width=0.45\linewidth]{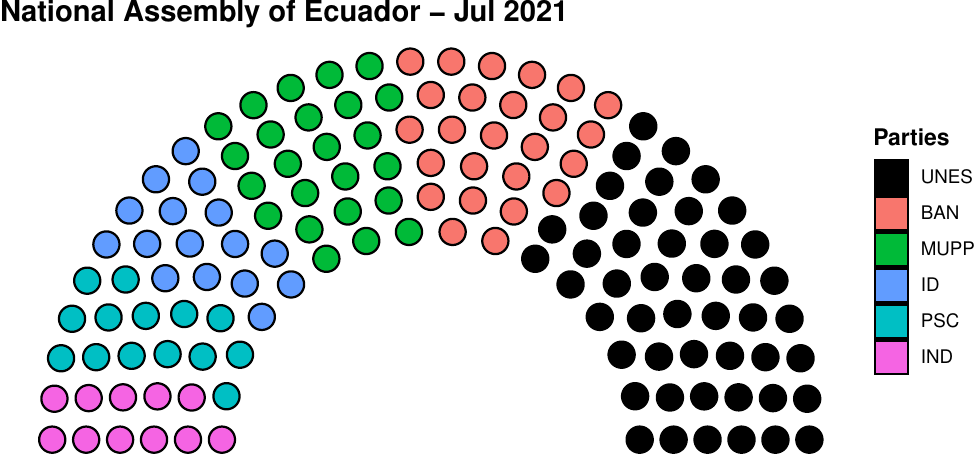} \\[0.5cm]
		
		\includegraphics[width=0.45\linewidth]{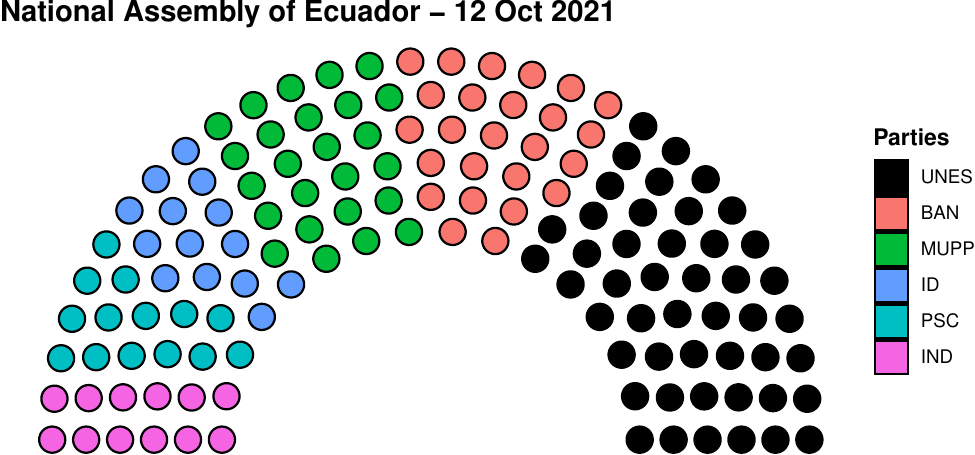} &
		
		\includegraphics[width=0.45\linewidth]{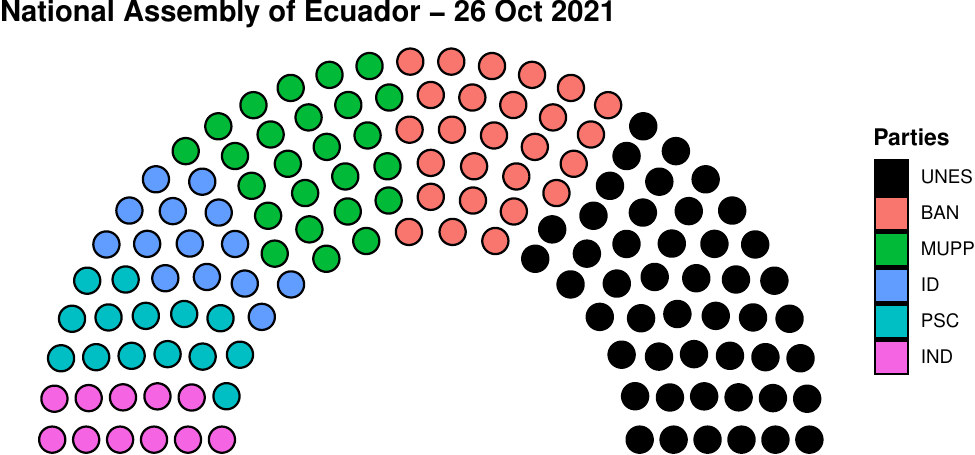} \\[0.5cm]
		
		\includegraphics[width=0.45\linewidth]{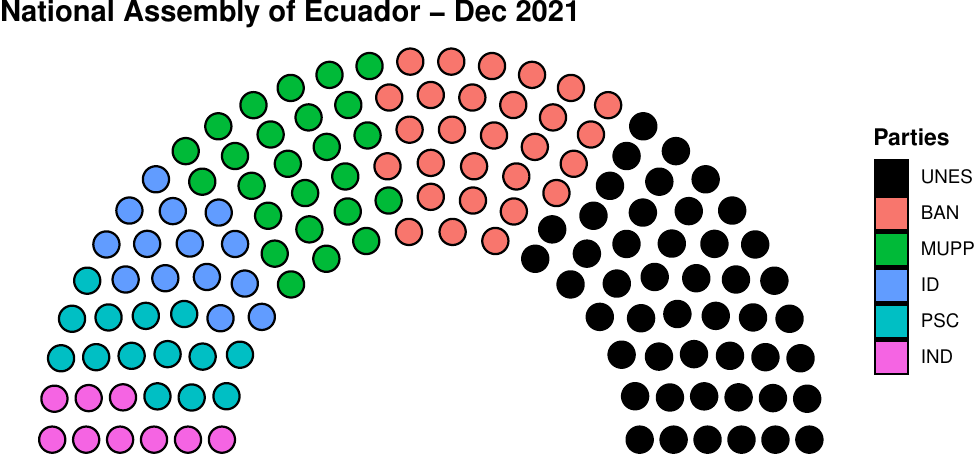} & \\
	\end{tabular}
	\end{center}
\end{figure}

In the seven months of operation of the National Assembly of Ecuador, there have been changes in the structure of its legislative benches. These changes are shown in Table \ref{Tb02} and Figure \ref{Fig10} and correspond to June 2021, July 2021\footnote{\url{https://www.primicias.ec/noticias/politica/bancadas-pierden-miembros-votos-asamblea/}, last accessed 23/12(2021.}, 12 October 2021\footnote{\url{https://www.primicias.ec/noticias/politica/union-unes-pachakutik-debilidad-legislativa-gobierno/}, last accessed 23/12/2021
}, 26 October 2021\footnote{\url{https://www.primicias.ec/noticias/politica/posible-destitucion-lasso-apoyo-asamblea/}, last accessed 23/12/2021}, and December 2021\footnote{\url{https://www.primicias.ec/noticias/politica/ruptura-pachakutik-capitulo-bancadas-desgranads/}, last accessed 23/12/2021
}.

\begin{table}
	\begin{center}
		\caption{Changes in the legislative benches in the National Assembly of Ecuador.}
		\label{Tb02}
		\small{
			\begin{tabular}{ r|| c|c | c | c | c  }
				\hline
				Benches & Jun 21 & Jul 21 & 12 Oct 21 & 26 Oct 21 & Dec 21 \\ \hline\hline
				UNES    & 48     &   47   & 47        & 47        & 47     \\
				MUPP    & 25     &   24   & 25        & 25        & 25     \\
				BAN     & 25     &   25   & 25        & 26        & 28     \\
				ID      & 16     &   16   & 14        & 14        & 14     \\
				PSC     & 14     &   14   & 14        & 14        & 14     \\
				IND     & 9      &   11   & 12        & 11        & 9      \\ \hline
			\end{tabular}}	
	\end{center}
\end{table}

Given that the National Assembly of Ecuador in May 2021 was structured in political parties and as of June 2021 it was structured in legislative benches, we divided the analysis for May 2021 and a comparative analysis from June to December 2021. 

We consider a decision procedure based on the plurality rule. Therefore, the games we will consider will be weighted majority games in partition function form as defined in Definition \ref{def:SGPFF}. The weights correspond first to the original National Assembly of Ecuador at May 2021 and subsequently those corresponding to the different legislative benches made, from June 2021 to December 2021. It should be kept in mind when defining the winning embedded coalitions that, in case of ties between coalitions with the highest overall weight, a tie-breaking rule should be used to determine the winner of an Assembly. One proposal is to count the number of votes in the elections and this criterion is the one we are going to use for May 2021. This tie-breaking rule was already used in \cite{Alonso2017} and in \cite{Arevalo2020}. In the cases corresponding to June, October and December 2021 this tie-breaking rule is not applicable since the legislative benches do not coincide exactly with the parties that presented themselves. So, one possibility is to consider that in case of a tie of two coalitions, both are winners, as proposed in \cite{Brink2021}. However, we will use the criterion that, in the event of a tie, none of the coalitions is considered the winner. This criterion is used in many voting systems: in the event that there is no single most voted option, the voting has to be repeated. In the case of the National Assembly of Ecuador there are few ties and either of the above two options for breaking ties lead to very similar results.

\subsection{Original formation of the National Assembly of Ecuador: May 2021}

First, we analyze the initial situation of the National Assembly of Ecuador in May 2021. In the case of ties, the vote received by each of the parties in the February 2021 general elections is used. In Table \ref{Tb06}, we observe the 34 minimal winning embedded coalitions for the National Assembly of Ecuador in May 2021. Table \ref{Tb17} shows the two ties found in determining the winning embedded coalitions. In this table, the columns show the number of seats of the different coalitions that make up the partition. As mentioned above, the active coalition for these partitions will be the one that, having the highest number of assembly members, has also obtained the highest number of votes in the February 2021 general elections.

\begin{table}
	\begin{center}
		\caption{Minimal winning embedded coalitions - National Assembly of Ecuador in May 2021.}
			\label{Tb06}
		\small{
			\begin{tabular}{ l  l }
				\hline
				Active coalition         & Partition                                      \\ \hline\hline
				\{UNES\}                 & \{ID, PSC, CREO\}, \{MUPP, IND\}, \{UNES\}     \\
				\{UNES\}                 & \{ID, CREO, IND\}, \{MUPP, PSC\}, \{UNES\}     \\
				\{UNES\}                 & \{PSC, CREO, IND\}, \{MUPP, ID\}, \{UNES\}     \\
				\{UNES\}                 & \{MUPP, CREO\}, \{ID, IND\}, \{UNES\}, \{PSC\} \\
				\{UNES\}                 & \{MUPP, CREO\}, \{ID, PSC\}, \{UNES\}, \{IND\} \\
				\{UNES\}                 & \{MUPP, CREO\}, \{PSC, IND\}, \{UNES\}, \{ID\} \\
				\{UNES\}                 & \{ID, IND\}, \{PSC, CREO\}, \{UNES\}, \{MUPP\} \\
				\{UNES\}                 & \{ID, PSC\}, \{CREO, IND\}, \{UNES\}, \{MUPP\} \\
				\{UNES\}                 & \{ID, CREO\}, \{PSC, IND\}, \{UNES\}, \{MUPP\} \\
				\{UNES, MUPP\}           & \{ID, PSC, CREO, IND\}, \{UNES, MUPP\}         \\
				\{UNES, ID\}             & \{MUPP, PSC, IND\}, \{UNES, ID\}, \{CREO\}     \\
				\{UNES, ID\}             & \{MUPP, CREO, IND\}, \{UNES, ID\}, \{PSC\}     \\
				\{UNES, ID\}             & \{MUPP, PSC, CREO\}, \{UNES, ID\}, \{IND\}     \\
				\{UNES, PSC\}            & \{MUPP, ID, IND\}, \{UNES, PSC\}, \{CREO\}     \\
				\{UNES, PSC\}            & \{MUPP, ID, CREO \}, \{UNES, PSC\}, \{IND\}    \\
				\{UNES, PSC\}            & \{MUPP, CREO, IND\}, \{UNES, PSC\}, \{ID\}     \\
				\{UNES, CREO\}           & \{MUPP, ID, IND\}, \{UNES, CREO\}, \{PSC\}     \\
				\{UNES, CREO\}           & \{ID, PSC, IND\}, \{UNES, CREO\}, \{MUPP\}     \\
				\{UNES, CREO\}           & \{MUPP, PSC, IND\}, \{UNES, CREO\}, \{ID\}     \\
				\{UNES, IND\}            & \{MUPP, PSC, CREO\}, \{UNES, IND\}, \{ID\}     \\
				\{UNES, IND\}            & \{MUPP, ID, CREO\}, \{UNES, IND\}, \{PSC\}     \\
				\{UNES, CREO, IND\}      & \{UNES, CREO, IND\}, \{MUPP, ID, PSC\}         \\
				\{MUPP, ID, PSC\}        & \{MUPP, ID, PSC\}, \{UNES, CREO\},  \{IND\}    \\
				\{MUPP, ID, PSC\}        & \{MUPP, ID, PSC\}, \{UNES, IND\}, \{CREO\}     \\
				\{MUPP, ID, PSC\}        & \{MUPP, ID, PSC\}, \{CREO, IND\}, \{UNES\}     \\
				\{MUPP, ID, CREO\}       & \{MUPP, ID, CREO\}, \{PSC, IND\}, \{UNES\}     \\
				\{MUPP, PSC, CREO\}      & \{MUPP, PSC, CREO\}, \{ID, IND\}, \{UNES\}     \\
				\{MUPP, ID, IND\}        & \{MUPP, ID, IND\}, \{PSC, CREO\}, \{UNES\}     \\
				\{MUPP, PSC, IND\}       & \{MUPP, PSC, IND\}, \{ID, CREO\}, \{UNES\}     \\
				\{MUPP, CREO, IND\}      & \{MUPP, CREO, IND\}, \{ID, PSC\}, \{UNES\}     \\
				\{ID, PSC, IND\}         & \{ID, PSC, IND\}, \{MUPP, CREO\}, \{UNES\}     \\
				\{MUPP, PSC, CREO, IND\} & \{MUPP, PSC, CREO, IND\}, \{UNES, ID\}         \\
				\{MUPP, ID, CREO, IND\}  & \{MUPP, ID, CREO, IND\}, \{UNES, PSC\}         \\
				\{ID, PSC, CREO, IND\}   & \{ID, PSC, CREO, IND\}, \{UNES\}, \{MUPP\}     \\ \hline
			\end{tabular}}	
	\end{center}
\end{table}

\begin{table}
	\centering
	\caption{Partitions that result in ties - National Assembly of Ecuador in May 2021.}
		\label{Tb17}
	\small{
	\begin{tabular}{ r|| c | c | c | c  }
		\hline
		Partition                                      & Seats P1 & Seats P2 & Seats P3 & Seats P4 \\ \hline\hline
		\{ID, PSC, IND\}, \{MUPP, CREO\}, \{UNES\}     & 49       & 39       & 49       & --       \\
		\{ID, PSC, IND\}, \{UNES\}, \{MUPP\}, \{CREO\} & 49       & 49       & 27       & 12       \\ \hline
	\end{tabular}}
\end{table}

In the initial situation of the National Assembly of Ecuador in May 2021, we observe that UNES is present in 22 of the 34 active coalitions. Therefore, UNES is the party with the most power in the National Assembly of Ecuador in May 2021. UNES is also the only party capable of having active coalitions of cardinality one (9 of 34).  Likewise, UNES is present in all active coalitions of cardinality two (12). The party of the president of Ecuador CREO is present in 10 of the 34 active coalitions.

Table \ref{Tb07} presents the results of the calculation of different power indices in partition function form (plurality rule) in the National Assembly of Ecuador in May 2021.  UNES is the party with the most power in National Assembly of Ecuador in May 2021. Note that, although MUPP has more assembly members than ID and PSC, according to the Deegan-Packel power index, ID and PSC have more power than MUPP. This is because ID and PSC are present in 3 of the 12 active coalitions of cardinality two. While MUPP is only present in 1 of these 12 active coalitions. MUPP, ID, and PSC are each present in 11 of the 34 minimal winning embedded coalitions. Of the four proposed power indexes, the Public Good power index gives the least power to UNES, but it still has the most power. This is because only the cardinality of the set of minimal winning embedded coalitions to which each player belongs is taken into account. Recall that UNES is present in 22 out of 34 minimal winning embedded coalitions, while the rest of the legislative benches are present in 10 or 11 of them.

\begin{table}
	\centering
	\caption{Power indices in partition function form - National Assembly of Ecuador in May 2021.}
		\label{Tb07}
	\small{
	\begin{tabular}{r|| c | c | c | c | c | c  }
		\hline
		 Power indices & UNES   & MUPP   & ID     & PSC    & CREO   & IND    \\ \hline\hline
		         $\DP$ & 0.4510 & 0.1078 & 0.1176 & 0.1176 & 0.1054 & 0.1005 \\
		         $\PG$ & 0.2933 & 0.1467 & 0.1467 & 0.1467 & 0.1333 & 0.1333 \\
		 $\mathsf{CM}$ & 0.5496 & 0.1415 & 0.0944 & 0.0944 & 0.0572 & 0.0630 \\
		$\mathsf{HCM}$ & 0.5334 & 0.1470 & 0.0980 & 0.0980 & 0.0594 & 0.0643 \\ \hline
	\end{tabular}}
\end{table}

\subsection{Formation of the National Assembly of Ecuador: from Jun 2021 to Dec 2021}

Table \ref{Tb08} shows the 37 minimal winning embedded coalitions for the National Assembly of Ecuador in June 2021. Let us note that there are five ties in determining the winning embedded coalitions (see Table \ref{Tb27}). As mentioned above, there is no winning embedded coalition in any of these partitions. UNES is present in 28 of the 37 minimal embedded coalitions and it is the only legislative bench in active coalitions of cardinality one. Note that, although BAN and MUPP have more assembly members than PSC and ID, the latter appear in more minimal winning embedded coalitions (9 and 11, respectively) than the former (8 each).

\begin{center}
\small{
		\begin{longtable}{ll}
		\caption{Minimal winning embedded coalitions - National Assembly of Ecuador in June 2021.}
		\label{Tb08}\\
		\hline
		Active coalition & Partition \\
		\hline
		\hline
		\endfirsthead
		\caption[]{(continued)}\\
		\hline
		Active coalition & Partition \\
		\hline
		\hline
		\endhead
		
		\{UNES\} & \{MUPP, IND\}, \{BAN, PSC\}, \{UNES\}, \{ID\} \\
		\{UNES\} & \{MUPP, IND\}, \{BAN, ID\}, \{UNES\}, \{PSC\} \\
		\{UNES\} & \{MUPP, IND\}, \{ID, PSC\}, \{UNES\}, \{BAN\} \\
		\{UNES\} & \{MUPP, ID\}, \{BAN, IND\}, \{UNES\}, \{PSC\} \\
		\{UNES\} & \{MUPP, ID\}, \{BAN, PSC\}, \{UNES\}, \{IND\} \\
		\{UNES\} & \{MUPP, ID\}, \{PSC, IND\}, \{UNES\}, \{BAN\} \\
		\{UNES\} & \{MUPP, PSC\}, \{BAN, IND\}, \{UNES\}, \{ID\} \\
		\{UNES\} & \{MUPP, PSC\}, \{BAN, ID\}, \{UNES\}, \{IND\} \\
		\{UNES\} & \{MUPP, PSC\}, \{ID, IND\}, \{UNES\}, \{BAN\} \\
		\{UNES\} & \{BAN, IND\}, \{ID, PSC\}, \{UNES\}, \{MUPP\} \\
		\{UNES\} & \{BAN, ID\}, \{PSC, IND\}, \{UNES\}, \{MUPP\} \\
		\{UNES\} & \{BAN, PSC\}, \{ID, IND\}, \{UNES\}, \{MUPP\} \\
		\{UNES\} & \{ID, PSC, IND\}, \{UNES\}, \{MUPP\}, \{BAN\} \\
		
		\{UNES, BAN\} & \{MUPP, ID, PSC, IND\}, \{UNES, BAN\} \\
		\{UNES, MUPP\} & \{BAN, ID, PSC, IND\}, \{UNES, MUPP\} \\
		\{UNES, PSC\} & \{MUPP, BAN, IND\}, \{UNES, PSC\}, \{ID\} \\
		\{UNES, ID\} & \{MUPP, BAN, IND\}, \{UNES, ID\}, \{PSC\} \\
		\{UNES, PSC\} & \{MUPP, ID, IND\}, \{UNES, PSC\}, \{BAN\} \\
		\{UNES, IND\} & \{UNES, IND\}, \{MUPP, BAN\}, \{ID, PSC\} \\
		\{UNES, ID\} & \{UNES, ID\}, \{MUPP, BAN\}, \{PSC, IND\} \\
		\{UNES, PSC\} & \{UNES, PSC\},   \{MUPP, BAN\}, \{ID, IND\} \\
		\{UNES, ID\} & \{BAN, PSC, IND\}, \{UNES, ID\}, \{MUPP\} \\
		\{UNES, ID\} & \{MUPP, PSC, IND\}, \{UNES, ID\}, \{BAN\} \\
		\{UNES, PSC\} & \{BAN, ID, IND\}, \{UNES, PSC\}, \{MUPP\} \\
		\{UNES, IND\} & \{MUPP, ID, PSC\}, \{UNES, IND\}, \{BAN\} \\
		\{UNES, IND\} & \{BAN, ID, PSC\}, \{UNES, IND\}, \{MUPP\} \\
		\{MUPP, BAN\} & \{ID, PSC, IND\}, \{MUPP, BAN\}, \{UNES\}\\
		
		\{UNES, ID, IND\} & \{UNES, ID, IND\}, \{MUPP, BAN, PSC\} \\
		\{UNES, PSC, IND\} & \{UNES, PSC, IND\},\{MUPP, BAN, ID\} \\
		\{MUPP, BAN, PSC\} & \{MUPP, BAN, PSC\}, \{UNES, IND\}, \{ID\} \\
		\{MUPP, BAN, ID\} & \{MUPP, BAN, ID\}, \{UNES, PSC\}, \{IND\} \\
		\{MUPP, BAN, ID\} & \{MUPP, BAN, ID\}, \{UNES, IND\}, \{PSC\} \\
		\{MUPP, ID, IND\} &  \{MUPP, ID, IND\}, \{BAN, PSC\}, \{UNES\} \\
		\{MUPP, ID, PSC\} & \{MUPP, ID, PSC\}, \{BAN, IND\}, \{UNES\} \\
		\{BAN, ID, IND\} & \{BAN, ID, IND\}, \{MUPP, PSC\}, \{UNES\} \\
		\{BAN, ID, PSC\} & \{BAN, ID, PSC\}, \{MUPP, IND\}, \{UNES\} \\
		
		\{MUPP, BAN, PSC, IND\} & \{MUPP, BAN, PSC, IND\}, \{UNES, ID\}\\
		
		\hline
	\end{longtable}
}
\end{center}

\begin{table}
	\centering
	\caption{Partitions that result in ties - National Assembly of Ecuador in June 2021.}
		\label{Tb27}
	\small{
	\begin{tabular}{ r||  c | c | c | c   }
		\hline
		Partition                                     & Seats P1 & Seats P2 & Seats P3 & Seats P4 \\ \hline\hline
		\{BAN, MUPP, PSC\}, \{UNES, ID\}, \{IND\}     & 64       & 64       & 9        & --       \\
		\{BAN, PSC, IND\}, \{MUPP, ID\}, \{UNES\}     & 48       & 41       & 48       & --       \\
		\{MUPP, PSC, IND\}, \{BAN, ID\}, \{UNES\}     & 48       & 41       & 48       & --       \\
		\{BAN, PSC, IND\}, \{UNES\}, \{MUPP\}, \{ID\} & 48       & 48       & 25       & 16       \\
		\{MUPP, PSC, IND\}, \{UNES\}, \{BAN\}, \{ID\} & 48       & 48       & 25       & 16       \\ \hline
	\end{tabular}}
\end{table}

We could observe changes in the minimal winning embedded coalitions for each game from June to December 2021 in the National Assembly of Ecuador. In July 2021, two of the the partitions that resulted in ties in June 2021 become minimal winning embedded coalitions, namely \{\{MUPP, PSC, IND\}, \{BAN, ID\}, \{UNES\}\} with active coalition \{MUPP,PSC, IND\}, and \{\{BAN, PSC, IND\}, \{MUPP, ID\}, \{UNES\}\} with active coalition \{BAN, PSC, IND\}. The cardinality of the set of minimal winning embedded coalitions is then 39. Table \ref{Tb37} shows the unique tie found in determining the winning coalitions in this case.

\begin{table}
	\centering
	\caption{Partitions that result in ties - National Assembly of Ecuador in July 2021.}
		\label{Tb37}
	\small{
	\begin{tabular}{r||  c | c | c    }
		\hline
		                                Partition & Seats P1 & Seats P2 & Seats P3 \\ \hline\hline
		\{BAN, MUPP, PSC\}, \{UNES, ID\}, \{IND\} & 63       & 63       & 11       \\ \hline
	\end{tabular}}
\end{table}

On 12 October 2021, the minimal winning embedded coalitions are 40. The winning embedded coalition \{\{UNES, ID, PSC\}, \{MUPP, BAN, IND\}\} with active coalition \{UNES, ID, PSC\} and the winning embedded coalition \{\{BAN, MUPP, PSC\}, \{UNES, ID\}, \{IND\}\} with active coalition \{BAN, MUPP, PSC\} (this partition resulted in ties in July 2021, see Table \ref{Tb37}) become minimal winning embedded coalitions. Moreover, with respect to July 2021, the minimal winning embedded coalitions associated with partitions \{\{MUPP, BAN, IND\}, \{UNES, PSC\},\{ID\}\} and \{\{MUPP, BAN, IND\}, \{UNES, ID\}, \{PSC\}\} change their active coalitions to \{MUPP, BAN, IND\} in both cases. Lastly, the embedded coalition \{\{MUPP, BAN, PSC, IND\}, \{UNES, ID\}\} with active coalition \{MUPP, BAN, PSC, IND\} ceases to be minimal since it contains to the winning embedded coalition \{\{MUPP, BAN, IND\}, \{PSC\}, \{UNES, ID\}\} with active coalition \{MUPP, BAN, IND\}.

The minimal winning embedded coalitions on 26 October 2021 are the same as on 12 October 2021. There are no ties in either case. Finally, we observe only one change in December 2021. The minimal winning embedded coalition \{\{BAN, ID, PSC\}, \{UNES, IND\}, \{MUPP\}\} with active coalition \{UNES, IND\} is out since its partition results in ties (see Table \ref{Tb47}). So, we have 39 minimal winning embedded coalitions in December 2021.

\begin{table}
	\centering
	\caption{Partitions that result in ties - National Assembly of Ecuador in December 2021.}
		\label{Tb47}
	\small{
	\begin{tabular}{ r||  c | c | c    }
		\hline
		Partition                                 & Seats P1 & Seats P2 & Seats P3 \\ \hline\hline
		\{BAN, ID, PSC\}, \{UNES, IND\}, \{MUPP\} & 56       & 56       & 25       \\ \hline
	\end{tabular}}
\end{table}

Table \ref{Tb09} shows the distribution of power of the different legislative benches in the National Assembly of Ecuador from June to December 2021. In the period of analysis, we observe that UNES is the legislative bench with the most power in the Assembly. The Deegan-Packel and Public Good power indices do not change the power assigned to MUPP and BAN in spite of changes observed in the structure of the legislative benches. Nevertheless, the Colomer-Martínez and $\mathsf{HCM}$ power indices show variations according to these changes.

\begin{table}
	\centering
	\caption{Power indices in partition function form - National Assembly of Ecuador.}
		\label{Tb09}
	\small{
	\begin{tabular}{r| r|| c | c | c | c | c }
		\hline
		                 Parties & Power indices  & Jun 21 & Jul 21 & 12 Oct 21 & 26 Oct 21 & Dec 21 \\ \hline\hline
		\multirow{4}{1 cm}{UNES} & $\DP$          & 0.5450 & 0.5171 & 0.4875    & 0.4875    & 0.4872 \\
		                         & $\PG$          & 0.3889 & 0.3590 & 0.3293    & 0.3293    & 0.3250 \\
		                         & $\mathsf{CM}$  & 0.6560 & 0.6179 & 0.5808    & 0.5821    & 0.5782 \\
		                         & $\mathsf{HCM}$ & 0.6346 & 0.5990 & 0.5618    & 0.5618    & 0.5547 \\ \hline
		\multirow{4}{1 cm}{MUPP} & $\DP$          & 0.0788 & 0.0833 & 0.1000    & 0.1000    & 0.1026 \\
		                         & $\PG$          & 0.1111 & 0.1154 & 0.1341    & 0.1341    & 0.1375 \\
		                         & $\mathsf{CM}$  & 0.0889 & 0.0943 & 0.1167    & 0.1164    & 0.1188 \\
		                         & $\mathsf{HCM}$ & 0.0944 & 0.0983 & 0.1217    & 0.1217    & 0.1248 \\ \hline
		 \multirow{4}{1 cm}{BAN} & $\DP$          & 0.0788 & 0.0833 & 0.1000    & 0.1000    & 0.1026 \\
		                         & $\PG$          & 0.1111 & 0.1154 & 0.1341    & 0.1341    & 0.1375 \\
		                         & $\mathsf{CM}$  & 0.0889 & 0.0973 & 0.1167    & 0.1201    & 0.1301 \\
		                         & $\mathsf{HCM}$ & 0.0944 & 0.1024 & 0.1217    & 0.1266    & 0.1398 \\ \hline
		  \multirow{4}{1 cm}{ID} & $\DP$          & 0.1171 & 0.1111 & 0.1042    & 0.1042    & 0.1068 \\
		                         & $\PG$          & 0.1528 & 0.1410 & 0.1341    & 0.1341    & 0.1375 \\
		                         & $\mathsf{CM}$  & 0.0791 & 0.0752 & 0.0645    & 0.0645    & 0.0660 \\
		                         & $\mathsf{HCM}$ & 0.0831 & 0.0801 & 0.0682    & 0.0682    & 0.0699 \\ \hline
		 \multirow{4}{1 cm}{PSC} & $\DP$          & 0.0968 & 0.1090 & 0.1042    & 0.1042    & 0.1068 \\
		                         & $\PG$          & 0.1250 & 0.1410 & 0.1341    & 0.1341    & 0.1375 \\
		                         & $\mathsf{CM}$  & 0.0546 & 0.0668 & 0.0645    & 0.0645    & 0.0660 \\
		                         & $\mathsf{HCM}$ & 0.0595 & 0.0701 & 0.0682    & 0.0682    & 0.0699 \\ \hline
		 \multirow{4}{1 cm}{IND} & $\DP$          & 0.0833 & 0.0962 & 0.1042    & 0.1042    & 0.0940 \\
		                         & $\PG$          & 0.1111 & 0.1282 & 0.1341    & 0.1341    & 0.1250 \\
		                         & $\mathsf{CM}$  & 0.0326 & 0.0485 & 0.0567    & 0.0525    & 0.0409 \\
		                         & $\mathsf{HCM}$ & 0.0340 & 0.0501 & 0.0584    & 0.0536    & 0.0409 \\ \hline
	\end{tabular}}
\end{table}

Next, we carry out some general comments. First, UNES is allocated with about half or more of the total power at almost all considered periods, with the exception of the Public Good power index. This may be due to the fact that, in the formation of winning embedded coalitions, the participation of UNES is preferable since it is the legislative bench with the highest number of votes. Second, with respect to the Deegan-Packel and Public Good power indices when the simple majority rule is used, these power indices do not show variation for the legislative benches in the period of analysis (see \citealp{Armijos2021}). This is because these power indices consider as valid the minimal winning coalitions, which do not change over the analyzed period. Whereas, when using the plurality rule, the minimal winning embedded coalitions change. Thus, the Deegan-Packel and Public Good power indices present changes in the distribution of power in the National Assembly of Ecuador. Third, the Colomer-Martínez and $\mathsf{HCM}$ power indexes, specifically defined for weighted majority games, better represent the changes made in the Assembly structure over the period of analysis. 

Finally, another notable result is that ID, using the Deegan-Packel and Public Good power indices, has greater power than BAN or MUPP. This is despite the fact that ID has fewer assembly members than BAN or MUPP. However, ID participates in a larger number of minimal winning embedded coalitions than BAN or MUPP. As noted, the Deegan-Packel and Public Good power indices distribute power equally among minimal winning embedded coalitions. However, the Colomer-Martínez and $\mathsf{HCM}$ power indices consider in their calculation the number of assembly members of each legislative bench. 

\section{Conclusions}\label{sec:con}

While the study of games in partition function form and simple games in partition function form is more abundant, it is less so for the case of weighted majority games in partition function form. In this paper we extend to this context two power indices axiomatically characterized in \cite{Armijos2021} in the context of classical weighted majority games, one of them introduced in \cite{Colomer1995} and the other defined in \citeauthor{Armijos2021}'s own work by combining the ideas of \cite{Colomer1995} and \cite{Holler1982}. Both extensions are illustrated by means of the case of the National Assembly of Ecuador in 2021 and the results obtained are compared with two power indices proposed and axiomatically characterized in \cite{Alonso2017} in the context of simple games in partition function form that turn out to be extensions of the Deegan-Packel power index (\citealp{Deegan1978}) and the Public Good power index (\citealp{Holler1982}) defined on the family of simple games.

In the case considered and under the adoption of the plurality rule, the two new power indices proposed turn out to be sensitive to the different relevant ingredients present, such as the resulting minimal winning embedded coalitions and the number of representatives of each party or political grouping. They are also capable of reacting to the measurement of power in a changing and complex situation such as the National Assembly of Ecuador. It is worthwhile to further investigate these new proposals of power indices for weighted majority games in partition function form in the future. Some directions are the study of their properties and the achievement of axiomatic characterization results, the exploration of combinatorial analysis techniques aimed at efficient computation and their application in other real life problems.

\section*{Acknowledgment}
This research has been supported by the ERDF (MINECO/AEI grants MTM2017-87197-C3-2-P and MTM2017-87197-C3-3-P) and by the Xunta de Galicia (Competitive Reference Group ED431C 2021/24 and ED431C 2020/03).
\bibliographystyle{apalike}
\bibliography{DP-bib}

\end{document}